\documentclass[aip,apl,reprint,amsmath,amssymb]{revtex4-1}

\usepackage[section]{placeins}
\usepackage{graphicx}

\begin{document}

\title{Fiber-pigtailed optical tweezer for single-atom trapping and
  single-photon generation}

\author{S. Garcia}
\author{D. Maxein}
\author{L. Hohmann}
\author{J. Reichel}
\author{R. Long}
\email{long@lkb.ens.fr}

\affiliation{Laboratoire Kastler Brossel, ENS, Universit\'e Pierre et Marie Curie-Paris 6, CNRS, 24 rue Lhomond, 75\,005 Paris, France}

\date{\today}

\begin{abstract}
  We demonstrate a miniature, fiber-coupled optical tweezer to trap a
  single atom. The same fiber is used to trap a single atom and to
  read out its fluorescence. To obtain a low background level, the
  tweezer light is chopped, and we measure the influence of the
  chopping frequency on the atom's lifetime. We use the single
  atom as a single-photon source at $780~$nm and measure the
  second-order correlation function of the emitted photons. Because of
  its miniature, robust, fiber-pigtailed design, this tweezer can be
  implemented in a broad range of experiments where single atoms are
  used as a resource.
\end{abstract}


\maketitle

Trapped single atoms are an enabling tool in quantum science and
technology, and are investigated for applications from quantum
information \cite{Schlosser01,Volz06,Wilk10,Isenhower10,Hofmann12} to
quantum sensing \cite{Thompson13}. On par with single ions, they also provide the best performance among all emitters for indistinguishable, narrow-band single photons
\cite{Eisaman11}. So far however, one of their major drawbacks has been the
size and complexity of single-atom sources \cite{Lounis05}.  Here we
demonstrate single-atom trapping and single-photon production with a
simple and practical, miniature optical tweezer. The device is fiber coupled, making it robust and simplifying its integration as part of a more complex experiment. It is also cheap to build and does not require cleanroom techniques. Chopping the dipole trap completely eliminates trap-induced light shifts and
broadening in the single-photon spectrum. In addition to single-photon
generation, this device substantially simplifies the production of
single atoms for applications in quantum information and quantum
optics.

\begin{figure}[h]
\includegraphics[width=7cm]{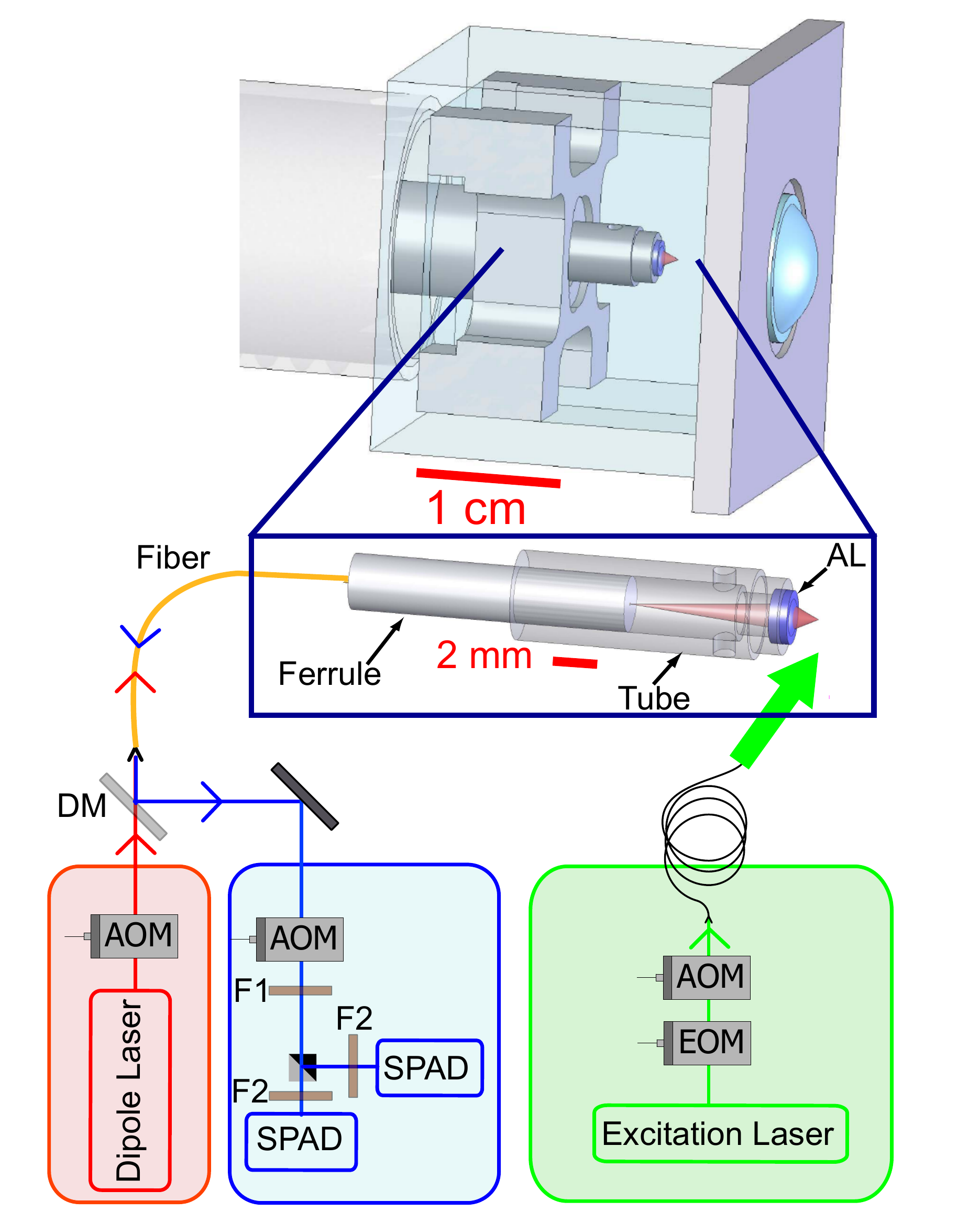}
\caption{\label{fig:setup} The fiber-optical tweezer: the aspheric
  lens (AL) is glued to the end of a ceramic tube, the fiber inside a
  ferrule is glued inside the bore of the tube leading to a $1.4~\mu$m
  waist at $1~$ mm from the lens. Laser setup: The dipole light and
  the fluorescence light are split by a dichroic mirror (DM). A custom
  filter (F1) and standard filters (F2) on the detection path to the
  single-photon avalanche diodes (SPAD) remove background dipole trap
  light. The excitation pulses for single photon generation are
  obtained by sending a continuous laser through an intensity EOM
  followed by an AOM.}
\end{figure}

A well-established technique to trap single atoms is the use of a
tightly confining far off-resonance optical trap. For a single,
red-detuned beam, atoms are attracted toward the beam focus due to the
dipole force. If this optical tweezer is confining enough, the
``collisional blockade'' effect efficiently eliminates states with
more than one atom, guaranteeing that no more than one atom is present
inside the trap \cite{Schlosser01}.  To enter the collisional blockade
regime, high numerical aperture optics are needed, making integration
and scalability challenging and costly.  On the other hand, it has
already been demonstrated that the trapping of single atoms is
possible with a single, commercially available aspheric lens, which is
placed inside the vacuum chamber \cite{Sortais07, Tey08,
Grunzweig10}. In those experiments however, the light enters and
leaves the vacuum as a free-space beam, requiring macroscopic lenses
and careful alignment, further compromising scalability. In contrast
to these setups, our approach relies on a miniature, fiber-pigtailed
device terminated with a small commercial aspheric lens and placed
entirely inside the vacuum chamber. By combining this device with the
technique of chopping the dipole light \cite{Chu86}, we are able to
use the same fiber for dipole light delivery and single-atom
fluorescence extraction with low background.

In our prototype setup, the aspheric lens (LightPath Technologies
Model 355200) is glued to the end of a machinable ceramic tube. The
fiber is glued inside a ceramic ferrule, which is inserted into the
bore of the ceramic tube (Fig. \ref{fig:setup}). We optimize the
position of the ferrule inside the tube to obtain the smallest
possible waist at the trapping wavelength ($1.4~\mu$m) before gluing
it inside the tube. The focus is at 1\,mm from the end face of the
lens.  We use $^{87}$Rb atoms with an emission wavelength of 780\,nm
(D2 line) and dipole trapping light with a wavelength close to
810\,nm. An inherent advantage of our design is that the trap position
and the collection focus coincide very well if the dipole and
fluorescence wavelength are not too far apart
\cite{Takamizawa06}. Along the optical axis, the two foci are distant
by $2\,\mu$m (about one third of the Rayleigh range), which ensures a
good collection efficiency without any further alignment.  This
prealigned system is placed inside a cubic, 25\,mm side length,
all-glass cell intended for spectroscopy (Hellma 704.000).  A simple
fiber feed-through \cite{Abraham98} is sufficient to bring the light
into and out of the vacuum chamber.  Loading of the dipole trap is
achieved by producing a cloud of laser-cooled atoms in a
magneto-optical trap (MOT) at the focus of the dipole beam. For the
MOT, we use three retro-reflected beams $1\,$mm in diameter, with two
of them crossing at an angle of $20\,^{\circ}$ to avoid clipping at
the lens.  The dipole light is chopped with an acousto-optical
modulator (AOM), single-photons being detected while the dipole trap
is off. This eliminates trap-induced light shifts as a source of
spectral broadening. It also avoids the generation of 780\,nm photons
by anti-Stokes Raman scattering of the trapping light inside the
fiber, which would obfuscate the single-atom fluorescence
\cite{Farahani99,Suh08}.

\begin{figure}
\includegraphics[width=9cm]{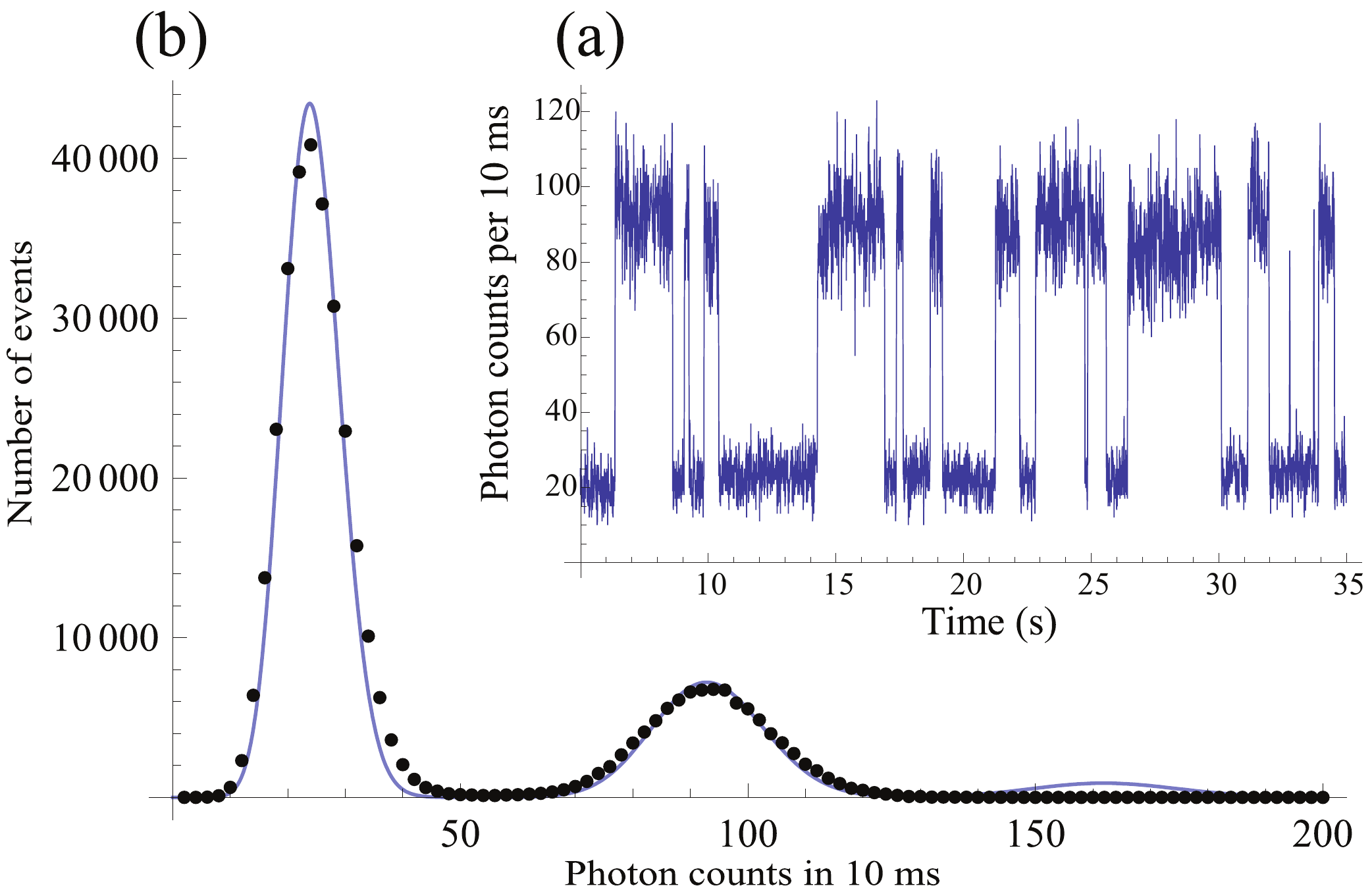}
\caption{\label{fig:Histo} \textbf{a)} Number of counted photons for $10\,$ms time bins showing step signal. \textbf{b)} Histogram of counted photons during $3850\,$s (black points)  with a fit by a compound Poisson law for 0, 1 and 2 atoms peaks (solid line).}
\end{figure}

In a first experiment, we keep the MOT light always on.  During the
dark phase (dipole light off), we count the fluorescence photons
emitted by an atom and collected via the lens and fiber. We separate
the fluorescence light from the dipole light with a dichroic mirror
(Semrock LPD01-785RU-25). To block background light of the dipole
laser, we use a custom interference filter centered at $780\,$nm with
a bandwidth of $0.3\,$nm and a transmission of about $90\,\%$,
followed by commercial interference filter (Semrock LL01-780-12.5).
Additionally, we use an acousto-optical modulator to open the
detection path only during the dark phase of the dipole laser, thus
preventing Raman photons from the fiber pigtail from reaching a
single-photon avalanche diode (SPAD). (These photons otherwise create
a background during the detection window, probably due to delayed
afterpulses of our SPAD\cite{Enderlein05}.)  The fluorescence light is
then coupled into a multimode fiber and detected by the SPAD
(Excelitas Technologies SPCM-AQRH-13-FC). With this setup, we observe
a background rate of about 100 counts$/$s (dark counts and remaining
Raman/afterpulse contribution) when the MOT is off, and 2100 cts$/$s
when the MOT is on.  This is low enough that in the presence of the
MOT beams, we can see a high-contrast two-level fluorescence signal,
jumping between photon count rates corresponding to zero or one atom
inside the trap (Fig.~\ref{fig:Histo}(a)). We do not observe any
two-atom events within our binning resolution of $10\,$ms, indicating
that the trap operates in the blockade regime where light-assisted
two-body collisions lead to a rapid loss of the two
atoms. Fig.~\ref{fig:Histo}(b) shows a histogram of counts recorded
during $3850\,$s. Fitting it with a compound Poisson law underscores
the complete absence of a two-atom peak, confirming the single atom
trapping ability of our trap.

\begin{figure}
\includegraphics[width=8cm]{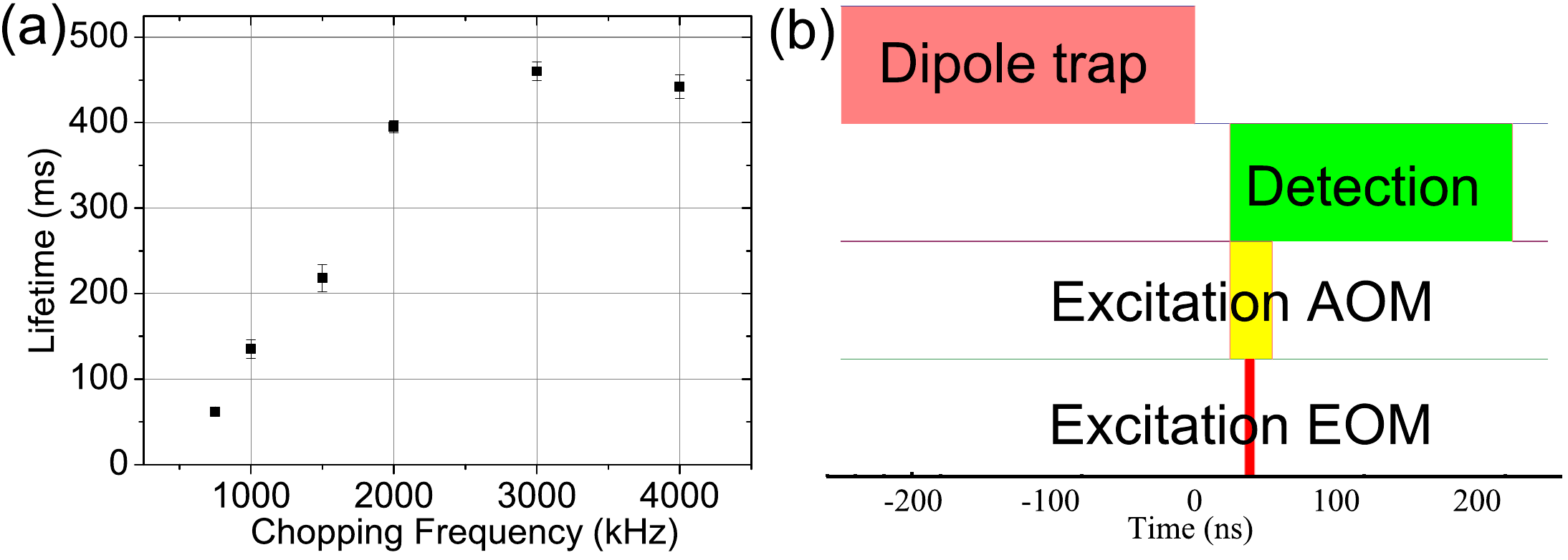}
\caption{\label{fig:Dipole_lifetime} \textbf{a)} Dipole trap lifetime
  in presence of the MOT versus chopping frequency, for linearly
  polarized dipole light.  \textbf{b)} Timing sequence. For the data
  in Figs.~\ref{fig:Histo} and \ref{fig:Dipole_lifetime}(a), we chop the
  dipole light and the detection path. For single photon generation,
  we switch off the MOT and apply resonant excitation pulses by using a CW
  laser passing through an intensity-EOM followed by an AOM.}
\end{figure}

We have investigated the influence of the trap chopping frequency on
the lifetime of the atom inside the trap in the presence of the MOT
beams. With a duty cycle of 50 $\%$, the time-averaged dipole light
power is $6.9\,$mW, corresponding to a trap depth of $2.8\,$mK.  To be
mainly limited by background gas collisions, we work in the weak
loading regime, where loss due to a second atom entering the trap is
negligible. When the dipole light is on, the transverse and
longitudinal trap frequencies are $167\,$kHz and $22\,$kHz,
respectively. For chopping frequencies larger than the trap
frequencies, we expect the atom to experience a time-averaged
potential in which it can stay trapped. In
Fig. \ref{fig:Dipole_lifetime}, we observe, with linearly polarized dipole light, a lifetime of about $400\,$ms for
frequencies above $2\,$MHz, limited by background gas collisions. For
frequencies below $2\,$MHz, the lifetime decreases with the chopping
frequency. Below $500\,$kHz, we could not detect any single-atom
trapping signal.  We notice that the lifetime of the atom is very
sensitive to the polarisation of the dipole light and is actually
improved by adding a small circular polarisation. This behavior could
be due to a residual magnetic field which is compensated by an
effective magnetic field generated by the circular polarisation of the
light, leading to a better cooling of the atoms inside the dipole trap
\cite{Mathur99}.

In the following, we use the single atoms as a single-photon source
\cite{Darquie05}.  For single photon generation, it is critical to
further reduce the background level. This is easily achieved by
switching off the MOT light after loading the dipole trap. To obtain a
single photon source, we use the following timing sequence.  The
chopping frequency is fixed at $2\,$MHz. During a loading phase, we
apply the MOT lasers and the magnetic field gradient until two
consecutive 10\,ms integration intervals of the fluorescence counts
are above a threshold which depends on the background and single-atom
count rates. This loading phase has an average duration of 1\,s. We
then turn the MOT lasers and the magnetic gradient off, leaving the
single atom inside the chopped dipole trap.  A stream of single
photons is produced from the trapped atom by resonantly exciting it
with a $\pi$-pulse each time the dipole light is off. The excitation
beam is tuned on the transition $\left|5^2 \mathrm{S}_{1/2},
  F=2\right\rangle \rightarrow \left|5^2 \mathrm{P}_{3/2},
  F'=3\right\rangle$ and aligned orthogonally to the trap axis (see
Fig. 1). The pulses are generated by sending a continuous laser
through an electro-optic intensity modulator (EOM, Jenoptik AM780HF).
The pulse duration is set to $\tau=3.5\,$ns in order to avoid double
excitation of the atom (excited state lifetime $\tau \ll 26\,$ns) and
excitation to other states due to Fourier-broadened linewidth ($\tau
\gg 1/(2 \Delta\omega) \simeq 1.9\,$ns, where $\hbar \Delta\omega$ is
the energy difference between the $F'=2$ and $F'=3$ excited
states). As the EOM's extinction ratio of 800 is not high enough to
completely avoid excitation during the off-phase, we add an AOM for
additional extinction. We pulse this AOM with a duration of $30\,$ns
(limited by the AOM response), centered on the EOM pulse. The
excitation beam waist is $50\,\mu$m at the position of the dipole
trap.  The polarization is $\sigma^{+}$ relative to a
$4\times10^{-4}\,$T quantization magnetic field applied in the laser
propagation direction.  In order to obtain a $\pi$-pulse, we adjust
the peak pulse power by observing the time distribution of the photons
detected by the SPAD after the pulse. This results in a peak power of
about $2\,$mW.

During the photon generation phase, the quantification field is
applied continuously and the timing sequence is as follows. After the
end of every dipole light pulse (i.e., every $500\,$ns), labeled
$t=0$, we open the detection path and the excitation AOM at
$t=25\,$ns. At $t=45\,$ ns, we apply the $3.5\,$ns $\pi$-pulse.  The
atom emits a single photon during the $200 \,$ ns detection window
with a calculated probability of $99.9 \,\%$.  While the dipole laser
is on, we also switch on a repumping laser, resonant on the transition
$\left|5^2 \mathrm{S}_{1/2}, F=1\right\rangle \rightarrow \left|5^2
  \mathrm{P}_{3/2}, F'=2\right\rangle$, to repump atoms that may have
fallen into the $F=1$ level.  The duration of the photon generation
phase is set to $2\,$ms, corresponding to 4000 excitation sequences,
after which another loading phase follows.  With such a short photon
generation phase, the atom remains trapped at the end of this phase
with high probability. This shortens the ensuing loading phase,
maximizing the average single-photon flux. During the photon
generation phase, the single-photon collection rate into the fiber is
about $13 500\,$ photons/s, which corresponds to a collection
efficiency of about $0.7 \,\%$.  On average, including the loading
phase, we obtain approximately $170$ single photons per second out of
the fiber. With the reasonable assumption that the single atom is a
Fourier-transform limited source, we obtain an average spectral
brightness of 28 photons/(s~MHz). This is better than state-of-the-art
diamond-based single-photon emitters \cite{Aharonovich11} and
comparable to some non-deterministic parametric down conversion
sources \cite{Haase09,Fekete13}, while keeping the advantage of atomic
sources in terms of indistinguishability of the emitted photons.

\begin{figure*}
\includegraphics[width=\textwidth]{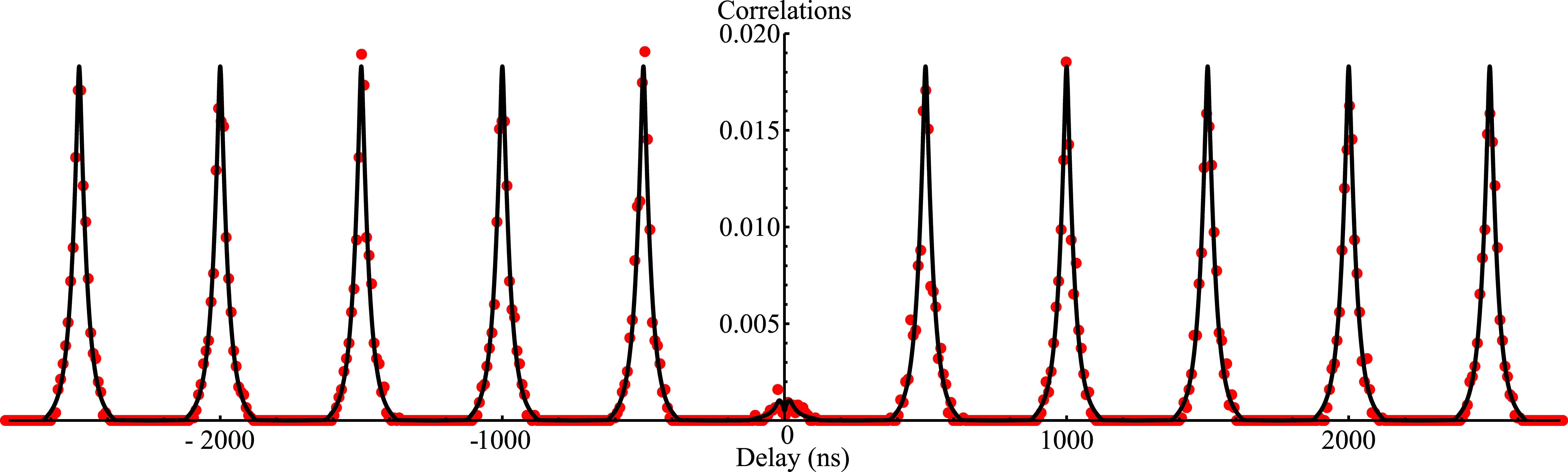}
\caption{\label{fig:g2} Second-order intensity
 correlations (normalized units). Red dots are experimental data with
  $8\,$ns binning. The black curve is an optical Bloch equation simulation including experimental noise. Near-zero coincidences around zero delay are the signature of a single photon source.}
\end{figure*}

To prove the single-photon characteristics of our source, we have
measured the second-order intensity correlations of the light field by
implementing a Hanbury-Brown and Twiss setup \cite{HBT56}.  The results are
presented in Fig.~\ref{fig:g2} without any background subtraction.
The peaks at multiples of $500\,$ns delay are due to correlations
between photons generated by different excitation pulses.  Half of the
relative area of the zero-delay peak, normalized to that of a
$500\,$ns delayed peak, gives approximately the probability to emit
two photons per excitation pulse \cite{McKeever04}.  We find a
probability of $(4.5 \pm 0.5) \%$ without background correction.
Subtracting the known contribution of SPAD dark count noise yields a
$(3\pm 0.6)\%$ probability of emitting two photons per pulse. This
limit is mainly due to the fact that the atom can be excited twice
during the detection window. The probability of double excitation is
$2\%$ during the pulse itself and $1\%$ during the time where the EOM
is off but the AOM still open. The latter contribution could be
avoided by using two EOMs in series.  We have simulated the excitation
using the optical Bloch equations, taking into account the EOM and AOM
pulse shapes as well as the detector dark counts. It fits the data
quite well (see Fig. 4) and gives a two-photon emission probability of
$3.3 \%$, close to the experimental value.

In conclusion, we have developed a miniature fiber-pigtailed optical
tweezer for single atom trapping and detection.
Though in this experiment the tweezer stays fixed, it is possible in
principle to move it within a vacuum chamber, providing a scanning cold
single-atom probe. An advantage of this approach is the possibility to
move the single atom without being limited by the transverse field of
view, enabling the delivery of single atoms inside an optical cavity
\cite{Chang09,Alton11,Thompson13} or the probing of surfaces at very short
distances \cite{Parazzoli12}. Further miniaturization is possible by
fabricating a lens directly on the fiber tip, reducing the volume of
the tweezer by another three orders of magnitude. We have also
demonstrated the use of the pigtailed tweezer as a single photon source
based on a single atomic emitter. Despite its low flux, we expect
this source to generate single photons with excellent
indistinguishability. This is due to the very good spatial mode
matching between single photons which are fiber-coupled by design,
and due to the fact that we excite the atoms when the dipole
trap is off, so that there is no broadening induced by light shifts.
These features make the fiber-pigtailed tweezer attractive for hybrid, cold
atom-surface science techniques as well as for complex quantum
engineering networks where single atoms are used as a resource.

\begin{acknowledgments}
  We acknowledge funding from \'{E}mergence-UPMC-2009 research program
  and from the EU STREP project QIBEC. D.~M. acknowledges a
  post-doctoral grant from the \'{E}mergence-UPMC-2009 research
  program. We thank I.~Saideh and M.~Ammar for contributions in the
  early stage of the experiment.
\end{acknowledgments}



%

\end{document}